\documentclass[12pt,preprint]{aastex}

\usepackage[utf8]{inputenc}
\usepackage{natbib}
\usepackage{epstopdf}
\usepackage{graphicx}

\newcommand\tstrut{\rule{0pt}{2.5ex}}

\slugcomment{}

\shorttitle{Porosity and Activation of Main Belt Comets}
\shortauthors{Haghighipour et al.}

\begin{document}

\title{Triggering the Activation of Main Belt Comets: The Effect of Porosity}

\author{N.\ Haghighipour}
\affil{Institute for Astronomy, University of Hawaii-Manoa, Honolulu, HI, 96825, USA }
\email{nader@ifa.hawaii.edu}

\author{T.\ I.\ Maindl}
\affil{Department of Astrophysics, University of Vienna, A-1180 Vienna, Austria}

\author{C. M. \ Sch\"{a}fer and O. J. Wandel}
\affil{Institut f\"{u}r Astronomie und Astrophysik, Eberhard Karls 
Universit\"{a}t T\"{u}bingen, 72076 T\"{u}bingen, Germany}

\begin{abstract}

It has been suggested that the comet-like activity of Main Belt Comets is due to the
sublimation of sub-surface water-ice that is exposed when these objects are impacted
by meter-sized bodies. We recently examined this scenario and showed that such impacts 
can in fact excavate ice and present a plausible mechanism for triggering the activation 
of MBCs (Haghighipour et al. 2016). However, because the purpose of that study was to prove 
the concept and identify the most viable ice-longevity model, the porosity of the object and 
the loss of ice due to the heat of impact were ignored. In this paper, we extend our impact
simulations to porous materials and account for the loss of ice due to an impact. We
show that for a porous MBC, impact craters are deeper, reaching to $\sim 15$ m implying that if 
the activation of MBCs is due to the sublimation of sub-surface ice, this ice has to be within 
the top 15 m of the object. Results also indicate that the loss of ice due to the heat of impact 
is negligible, and the re-accretion of ejected ice is small. The latter suggests that the activities 
of current MBCs are most probably from multiple impact sites. Our study also indicates that in order 
for sublimation from multiple sites to account for the observed activity of the currently known MBCs, 
the water content of MBCs (and their parent asteroids) needs to be larger than the values traditionally 
considered in models of terrestrial planet formation.

\end{abstract}


\section{Introduction}

With orbital and dynamical properties characteristic of asteroids, and tails similar 
to those of comets, Main-Belt Comets (MBCs) have attracted a great deal of interest since their 
identification as activated asteroids by Hsieh \& Jewitt in 2006. Most of the interest 
in these objects is due to the implication that their comet-like activity is the result of 
the sublimation of sub-surface volatiles, presumably water-ice. This, combined with results 
of dynamical studies that suggest MBCs are native to the asteroid
belt (see below), argues strongly in support of the idea that water-carrying planetesimals 
and planetary embryos from the outer part of the asteroid belt provided the majority of 
Earth's water during its formation.

At the time of this writing, 8 unambiguous MBCs were known\footnote{We call an MBC unambiguous if its
activation can only be explained by sublimation of volatiles.}.
Table 1 and figure 1 show these objects along with some of their physical and orbital properties.
As shown here, MBCs are km-sized bodies with orbits that are mainly
in the outer part of the asteroid belt. Dynamical studies by \citet{Hagh09,Hagh10}
and \citet{Hsieh16} strongly suggest that these objects are most 
probably fragments of larger asteroids, and were scattered to their current orbits through
interactions with giant planets. This scenario is also supported by the fact that three 
of these objects, namely, 133P/(7968) Elst-Pizzaro, 313P/Gibbs, and P/2012 T1 (PANSTARRS) 
are members of two asteroid families \citep{Nesvorny08,Hsieh13,Hsieh15}. We refer the reader 
to  \citet[][hereafter Paper I]{Hagh16} for a comprehensive review 
of the origin, dynamics, and activation of MBCs.
 
It has been suggested that the activity of MBCs is most likely due to
the sublimation of sub-surface water-ice that has been exposed through impacts of these objects 
with small, meter-sized bodies \citep{Hsieh04,Hsieh06}. We would like to note in addition to
collision, rotational disruption, for instance due to YORP effect \citep{Steckloff16,Graves17} 
or rapid rotation \citep{Sheppard15} has also been proposed as 
a mechanism to break up comets and activated asteroids, and expose their sub-surface volatiles.
In Paper I, we examined the collision-activation scenario of MBCs by modeling impacts using our SPH code. We showed that 
for a wide range of material strength and water content of MBCs, and for impact velocities typical of 
those in the asteroid belt, the impact craters are deep and large enough to expose water-ice.
Our results also indicated that the depths of impact craters, resulted 
from collisions of m-sized bodies with km-sized objects, are slightly larger than 10 m implying that 

\begin{itemize}
\item if the activity of MBCs is due to the sublimation of water-ice, ice must be buried no deeper than
approximately 15 m from the surface of the object, and 
\item the water content of MBCs (as well as those planetesimals and planetary embryos responsible 
for the delivery of water to the accretion zone of Earth) must be much higher than the 5\% water-mass 
fraction that is traditionally considered in models of terrestrial planet formation. 
\end{itemize}

While results of our simulations in Paper I clearly demonstrated that impacts of m-sized objects
can in fact excavate ice (and other volatiles) to trigger the activation of MBCs, they had
some limitations. Because in that paper, focus was placed on proving the concept and identifying
the most viable ice-longevity model that would be consistent with the sublimation-driven
activity of MBCs, the porosity of the target and the loss of ice due 
to the heat of impact were ignored. Porosity can play an important role as porous materials
have lower strength which causes impactors to penetrate deeper in the target,
and also affects the geometry and morphology of the impact crater. For instance,
while the results presented in paper I were consistent with the ice-longevity model proposed
by \citet{Schorghofer08} (this author suggests that a thin layer of solid material on the surface of an
asteroid would be sufficient to preserve water-ice for the age of the solar system),
if porosity causes impact craters to be deeper than 50 m, a competing theory by 
\citet{Prialnik09} for ice-longevity in asteroid belt may also be applicable. These authors 
suggest that water-ice inside an asteroid can sublimate during the evolution of the solar system 
causing the ice level to sink to depths below 50 m.

In this paper, we 
extend our simulations to include porous targets, and examine the degree to which porosity 
plays a role in the final depth and size of an impact crater, as well as the plausibility of
the two ice-longevity models. 
To consider porosity, we implemented in our SPH code the $P\!-\!\alpha$ porosity 
model of \citet{Jutzi08}, and simulated the impact between a m-sized body and a km-sized object 
for different impact velocities, impact angles, and water contents of an MBC.

The outline of our paper is as follows.
We continue in section 2 by explaining our computational method and the implementation of
porosity. In section 3, we present results of our impact simulations and present a comparison
between these results and those in Paper I. We conclude our study in sections 4 and 5 by 
discussing the implications of the results and presenting highlights of our findings.

\section{SPH Simulations and Initial Set-up}

To simulate impacts, we use a 3D SPH code developed by \citet{Schafer07,Schafer16} and 
\citet{Maindl13}. This code solves the continuity equation and the equation 
of the conservation of momentum in continuum mechanics. It also includes material strength and 
implements a full elasto-plastic  model \citep[see, e.g.,][]{Maindl13,Maindl14}.
We model any specific material using the Tillotson equation of state \citep{Tillotson62,Melosh96}.
Fracture and brittle failure are treated using the Grady-Kipp fragmentation prescription 
\citep{Grady93,Benz94,Benz99}. This prescription is based on flaws that 
are distributed in the material following a Weibull distribution with material-dependent parameters. 

The colliding bodies are discretized into mass packages (known as SPH particles) with each package 
carrying all physical properties (e.g.,  mass, momentum, energy) of the part of the solid body that 
it represents. As a result, depending on the type of the impactor or target material, particles may 
have different material parameters such as bulk and shear modulus and yield strength, or have different 
activation thresholds for the development of cracks. Each SPH particle moves as a point mass 
following the equation of motion. 

To include porosity, we implemented an extension of the so called $P\!-\!\alpha$ model by 
\citet{Herrmann69} as described by \citet{Jutzi09}. Conceptually, this model 
is based on dividing the change in the volume of a porous material into two parts; 
the pore-collapse of the porous material, and the compression of the matrix material.
These two parts are connected via a \emph{distention parameter} $\alpha$ defined as
\begin{equation}
\label{equationalpharhorhos}
\alpha = \frac{\rho_s}{\rho}\,.
\end{equation} 
In this equation, $\rho$ is the density of the porous material and $\rho_s$ is the density of the 
corresponding matrix material. Following \citet{Carroll72}, the pressure of the 
porous material $(P)\/$ can be expressed as a function of the distention parameter $(\alpha)$
and the pressure of the solid material $(P_s)$ as
\begin{equation}
\label{equationpalphaps}
P=\frac{1}{\alpha}P_s(\rho_s,E_s) = \frac{1}{\alpha}P_s(\alpha \rho, E)\,.
\end{equation}
Quantities $\rho_s$, $\rho$, $E_s$ and $E$ in equation (2) represent the density and internal 
energy of the solid and porous material, respectively. The internal energy corresponds to
the energy contained inside the system due to the thermodynamical state of its internal parts
excluding the kinetic energy of the object due to its bulk motion and its potential
energy due to an external force. We note that in equation (2), it has been 
assumed that the energy of the surface pores (i.e., the energy necessary to change 
the assembly of pores on the surface of an object) are negligible and, therefore, the energy of 
the porous material is equal to that of the solid material \citep[i.e. $E=E_s$,][]{Carroll72}. 
We use the Tillotson equation of state \citep{Tillotson62} to calculate the pressure as a 
function of $\rho$, $E$, and $\alpha$.

\section{Results of Impact Simulations}

We considered a similar set up as in Paper I and simulated the impact between a m-sized 
impactor and a km-sized target. Because we are interested in the effect of porosity, 
we carried out simulations for four different cases: a dry and non-porous target, 
a non-porous target with 50\% water-mass fraction of non-porous ice, a dry and porous target
with 50\% porosity (i.e., $\alpha=2$), and a 50\% porous target containing 50\% water-mass 
fraction of 50\% porous ice. We considered the impactor and the solid part of the target to be
basalt\footnote{Although MBCs are most probably carbonaceous chondrites (CC), at the moment,
no equation of state is known for CC material. Also, the purpose of this study is merely to
understand the significance of including porosity in impact simulations. Because the equation
state of basalt is well known, we, therefore, considered objects to be basaltic. A comprehensive
model of the impact of m-sized bodies with km-sized CC MBCs is currently in the works.}, 
and because the size of the impactor is much smaller than the target, we considered the impactor 
to be non-porous. The material parameters for basalt and ice used in the Tillotson equation of state, 
and the Weibull parameters for the flaw distributions are given in Table 2.

We resolved the combined system of the impactor and target into approximately 500,000 SPH particles. 
Because compared to the time of the influence of the gravitational force of the target body, the 
impact timescales are very short (the collision velocities are in the order of km/s 
whereas the MBCs' surface escape velocities are less than a few m/s, see Table 1), we simulated 
collisions without self-gravity \citep[see][for more details]{Maindl15}. 
To analyze the evolution of the system during each impact, we took 100 snapshots every 0.4\,ms. 
In between snapshots, time integration was continued with an adaptive step-size. 

We carried out simulations for impact velocities of 1.5, 2.5, 3.5, 4.4, and 5.3 km/s.
These values were chosen based on the study by \citet{Bottke94}, who showed that
for objects of 50 km and larger, impact velocities in the asteroid belt have a mean value of 
$\sim 5.3$ km/s with a most probable value at 4.4 km/s. We considered an abundance
of objects smaller than 50 km with similar orbital elements (e.g., semimajor axis,
eccentricity, inclination) in the asteroid belt (i.e., $e\lesssim 0.25\/$).
Given the small size of these objects, and therefore, their small gravitational interactions,
collisions between these bodies will occur with relative velocities much smaller than a few
km/s. Combining this assumption with results from \citet{Bottke94}, we considered a
range of impact velocities from 1.5 km/s to 5.3 km/s. The impact angles were chosen to 
be 0, $30^\circ$ and $45^\circ$.

Figure 2 shows snap shots of the final craters of two sets of simulations for an impact 
velocity of 4.4 km/s. 
The target is a mixture of 50\% porous basalt and 50\% porous water-ice and has a 50\% water-mass 
fraction. The left column shows the impact for a head-on collision and the right column shows 
the results for an impact angle of $\beta=30^\circ$. The orange color represents porous basalt and 
blue is porous water-ice. 
As expected, water-ice is exposed in the interior part of the impact crater and is also
scattered out due to the impact. Movies of these simulations can be found in the electronic 
supplementary material. 

A comparison between these results and those of non-porous simulations points to interesting 
differences. The most prominent difference is in the shape and morphology of craters. 
Figure 3 shows impact craters of simulations with porous (top) and non-porous (bottom) targets.
Both objects have a water-mass fraction of 50\%. The impact velocities in all simulations 
are 4.4 km/s. As shown here, 
craters in the porous targets are noticeably deeper and narrower, and extend in the direction 
of impact velocity. In contrast, the craters in non-porous targets are shallower and much wider. 
Figure 4 shows this more clearly and for all our simulations with different target material and 
different impact velocities. 

Figures 2 and 3 also show that craters form in a very short time and have irregular shapes.
This asymmetry in the shapes of the final craters seems to be in contrast with the works of \citet{Collins14} 
and \citet{Milbury15} who assumed that except for the most oblique cases, all impacts produce approximately 
radially symmetric craters. We believe that the reason for the quick formation of craters in our
simulations and their irregular shapes lies in the fact that the gravity of our targets (i.e., MBCs)
are negligible. Gravity is the main factor in forming final craters from transient
ones. In the absence of gravity, the plastic flows during the impact phase stop 
rather quickly after the impact. In our systems, the MBCs do not have much gravity and as a result, 
the craters are formed quickly and are solely strength-dominated. We refer the reader to \citet{Collins09} 
for crater formation in oblique impacts without gravity.

Because in porous targets, craters are irregularly shaped, we determined their depth
by directly measuring the distance between the lowest point of their crater to the surface
of the target. To determine the surface area of a crater, we followed the methodology presented in Paper I 
and calculated the area by fitting an ellipsoid to the crater. We refer the reader to sections 
3.1 and 3.3 of of Paper I for more details on the technical
aspects of our calculations. The increase in the penetration of the impactor in a porous target
can, then, be attributed to the fact that compared to non-porous objects, porous targets, especially those with 
mixture of water-ice, have lower material strength. As a result, when these objects are impacted,
the momentum and energy of the impact carry the impactor deeper in the target whereas in non-porous objects, 
the rapid compaction of the target at the impact site causes the energy of the impact to be transferred
laterally creating a less deep but wider crater. An important implication of the results shown by 
figure 4 is that crater depths are still smaller than 15 m suggesting the model by 
\citet{Schorghofer08} as the most viable ice-longevity model in the asteroid belt.

It has been suggested that the activity of an MBC, in addition to ice sublimation from the bottom 
and interior of an impact crater, may also be due to the sublimation of scattered ice that was
re-accreted on the surface of the MBC. To examine this scenario, we calculated the amount and
velocity of ejected ice after each collision. Figures 5 and 6 show the results. Figure 5 shows 
the amount of scattered ice in terms of the impact velocity and figure 6 separates this quantity
into groups based on the ejection velocity of scattered ices. The vertical axis in this figure
shows the accumulative mass of the ejected ice and the horizontal axis shows its velocity. Each
curve corresponds to a different impact velocity for both porous and non-porous targets.
As shown here, in all simulations, the ejection velocity of ice is larger than 20 m/s. An examination
of Table 1 indicates that this ejection velocity is almost 10 times greater than the largest 
escape velocity of the currently known MBCs implying that almost all ejected ice is lost and
there is basically no re-accretion. This strongly suggests that the activity of MBCs is most
likely due to ice sublimation from multiple impact sites.

We also studied the change in the porosity of the target due to an impact.
Figure 7 shows variations in the porosity 
of the targets of figure 2 during an impact. The color coding represents the value of the distention 
parameter $\alpha$ corresponding to the porosity of the target. Yellow represents 50\% porosity where 
$\alpha=2$ and black corresponds to no porosity where $\alpha=1$. The left column corresponds to a dry, 
porous target and the right column represents the same object with 50\% water content. As shown here, 
material on the surface of the impact crater is strongly compacted with the strongest compaction 
occurring at the bottom where the target becomes non-porous. As the shock of the impact propagates
throughout the object, the degree of compaction lessens at deeper 
distances suggesting that away from the impact site and well inside the object, the target maintains its 
original porosity. Our simulations show that the propagation of shocks do not cause the target to
disintegrate, and therefore, in addition to the maintaining its original porosity, the target maintains 
its original water content as well. The latter has important implications for the delivery of water to the 
accretion zone of Earth with water-carrying planetesimals and planetary embryos. As the orbits of these objects
evolve during their dynamical evolution and they reach the accretion zone of Earth, they are repeatedly
impacted by planetesimals and planetary embryos. However, as shown here and given the sizes of these objects, 
they can still maintain their water-ice deep inside until they are accreted by the still-forming Earth. 
Movies of the simulations of figure 7 can be found in the electronic supplementary material.

\section{Discussion}

We carried out an extensive analysis of the impact of a m-sized body with a km-sized MBC. 
We extended our previous simulations (Paper I), where objects were considered to 
be non-porous, to more realistic cases where the porosity of an MBC is taken into account. 
We carried out simulations for different values of impact velocities and impact angles, and
considered different water contents for the target. Results of simulations indicate 
that as expected, substantial amount of water-ice is exposed on the interior 
surface of impact craters providing a viable pathway for triggering activity of MBCs.
Results also indicated that the depth and size of craters 
increase for porous targets, however, the increase in depth is still within the regime ($<$ 15 m)
where the ice-longevity model by \citet{Schorghofer08} applies.

In addition to being more realistic, our new simulations advanced those in Paper I
by including vaporization due to the heat of impact. We treated
phase transition during ice vaporization at the time of the impact by using the 
Tillotson equation of state. Results are shown in figure 8.
As shown here and in agreement with the results obtained from observations, 
the amount of ice vaporized during an impact is very small. For instance, for the case
of 176P/(118401) LINEAR, the entire ice vaporization due to an impact is less than 5 tons
whereas the rate of ice-sublimation due to the activation of this body is $\sim 720$ kg/day.
Other MBCs sublimate about an order of magnitude higher per days. This finding suggests that
when modeling impacts as a way of excavating sub-surface ice to trigger activation of MBCs, 
vaporization due to impacts can be safely ignored.

As mentioned earlier, the combination of the high material strengths of our targets (see Table 2),
small sizes of our impactors, and very low surface gravity of MBCs points to the fact that our impacts 
and their final craters
are strength-dominated. This has strong implications when comparing our results with previous studies, 
in particular those of \citet{Richardson07}. These authors used the mathematical model developed by 
\citet{Holsapple93} and
presented a thorough study of many impact properties of comets. A comparison of our results with those
of these authors indicates that although our results are comparable with their findings within the order of magnitude, 
our crater diameters are smaller. This is not unexpected as our impacts involve asteroids which
naturally have higher dynamic material strengths \citep{Asphaug02} compared to soft targets such as comets 
\citet{Basilevsk16}. For instance,
our assumed porous, wet MBC material has an average density of 685 kg/m$^3$. With
an effective MBC diameters between 0.3 km and 4.0 km (see Table 1), the mean surface 
gravity of our targets range from 0.057 mm/s$^2$ to 0.38 mm/s$^2$, mostly lower 
than \citet{Richardson07}'s nominal value of 0.34 mm/s$^2$. Given the range of our 
impact velocities (1.5 km/s - 5.3  km/s), our crater diameters fall 
between 3 m and 13.3 m, which, quite understandably, are smaller than those presented by \citet{Richardson07}.
The crater diameters estimated by these authors range between 22 m to 26 m, and correspond to
considerably faster projectiles (10.2 km/s) impacting softer targets.

Our assessment of the amount of the re-accreted ice after an impact indicated that,
because of the low gravity of the target, except for cases where the impact 
velocities are very low, most scattered ice particles
are lost and are not accreted back. This finding is consistent with previous results
as reported in Paper I, and confirm that activation of MBCs must be due to ice sublimation 
from multiple impact sites.

In this study, we did not consider a regolith layer on the top of the target. We assumed a random
distribution for ice inclusions and considered ice to exist everywhere throughout the target
including its top surface. Although inclusion of a regolith layer might have resulted in
craters with slightly smaller depths, the scattered fragments of the regolith layer could
impact other parts of the target and expose ice in other sites causing underlying
ice to be exposed in a larger area. The latter may compensate for smaller ice re-accretion and smaller
ice exposure in the main impact crater. This scenario is currently being investigated.

\section{Concluding Remarks}

We close this study by presenting highlights of our findings.

\begin {itemize}
\item Impacts of small bodies presents a viable mechanism for exposing sub-surface volatiles including
water-ice to trigger sublimation-driven activity of MBCs.

\item The loss of ice due to the heat of impact is negligible.

\item Most of the ejected ice particles are lost and not re-accreted.

\item A comparison between ice sublimation from impact craters obtained from our simulations with results 
of observations suggests that activity of the current MBCs are most probably from multiple impact sites.

\item Results of simulations suggest that the water content of MBCs and those of their parent asteroids 
needs to be larger than those traditionally considered in the models of terrestrial planet formation
so that the sublimation of the exposed water-ice can
account for the rate of sublimation obtained from observations of these objects.

\item If the activation of MBCs is due to the sublimation of sub-surface water-ice, this ice 
must be buried within the top 15 m. This result points to  
the model of ice-longevity by \citet{Schorghofer08} as the most viable
model for the retention of water-ice in the asteroid belt. This author suggested that a small
layer of regolith on the outer surface of an asteroid can allow the body to maintain its 
sub-surface water ice for the age of the solar system. 

\end{itemize}

\acknowledgements
NH acknowledges support from NASA PAST program under grant NNX14AJ38G.
T.I.M. acknowledges support from FWF Austrian Science Fund, project S11603-N16.
Most of the numerical simulations in this study were performed on the bwFor-CLuster BinAC, 
supported by the state of Baden-Württemberg through bwHPC, and the German Research Foundation 
(DFG) through grant INST 39/963-1 FUGG.

\clearpage

\begin{figure*}
\centerline{\includegraphics[width=5in]{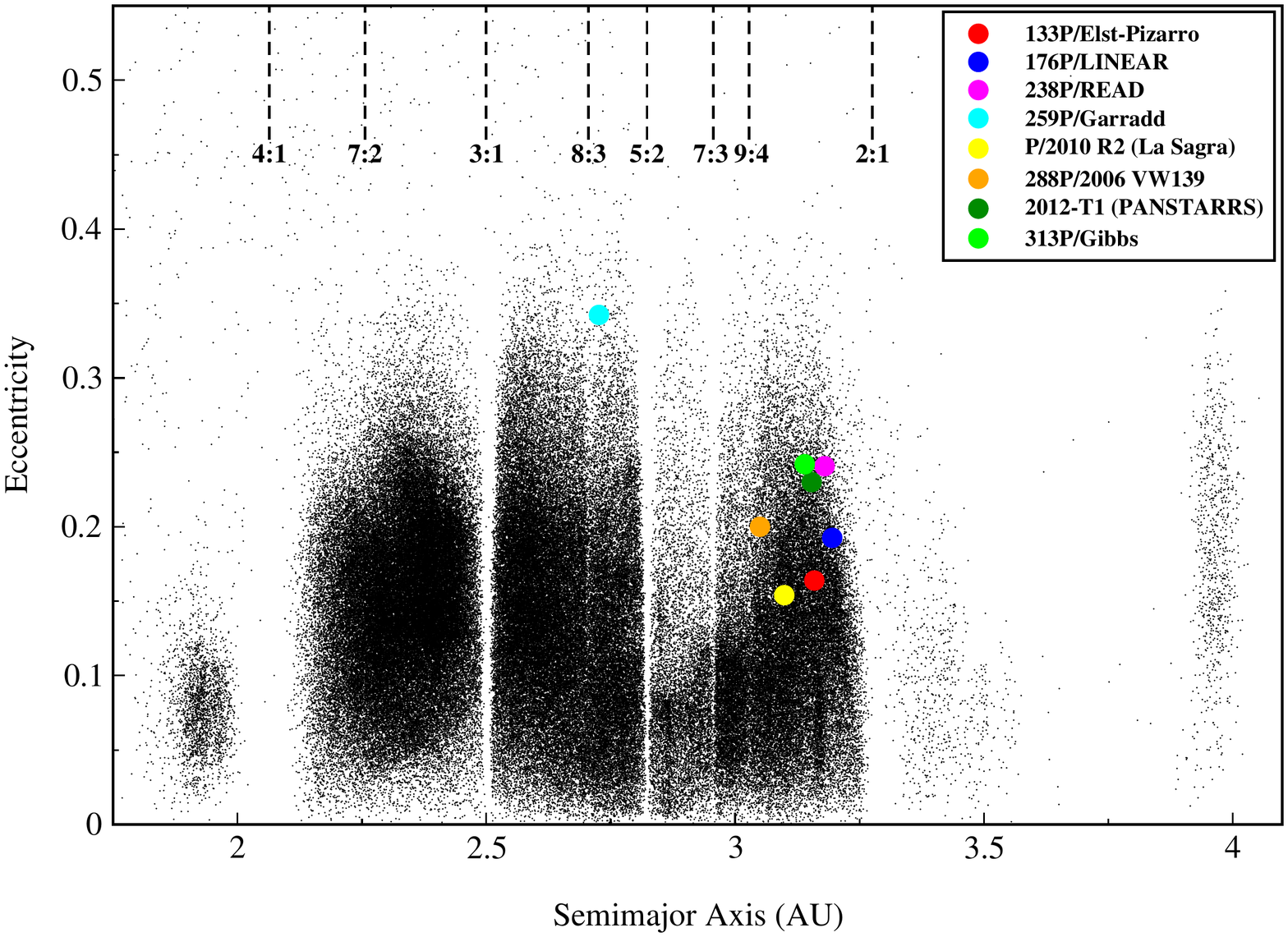}}
\caption{Locations of the currently known MBCs in the asteroid belt. 
The background shows all asteroids and the positions of
mean-motion resonances with Jupiter. }
\label{figure1}
\end{figure*}

\clearpage

\begin{figure}
\vskip -20pt
\center
\includegraphics[scale=0.9]{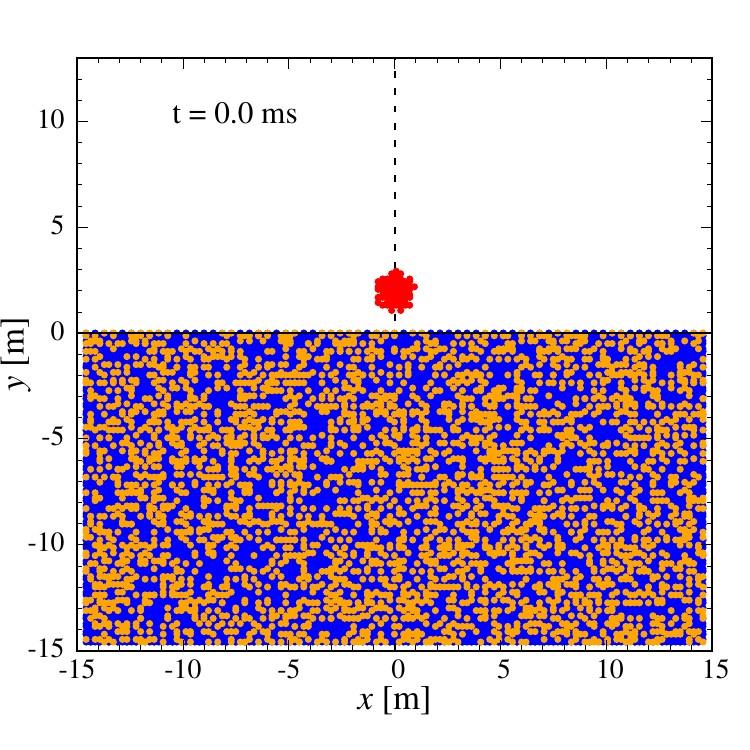}
\includegraphics[scale=0.9]{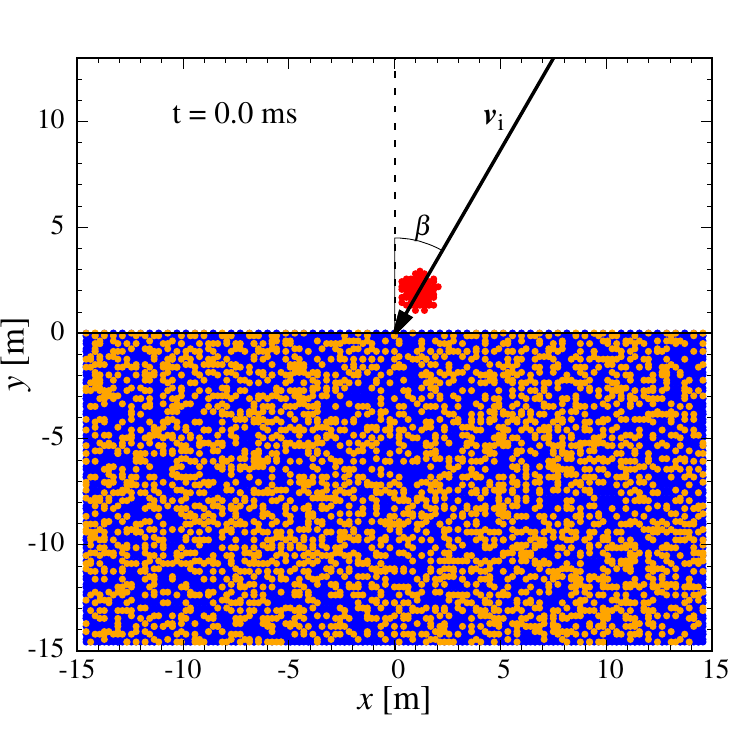}
\vskip -15pt
\includegraphics[scale=0.9]{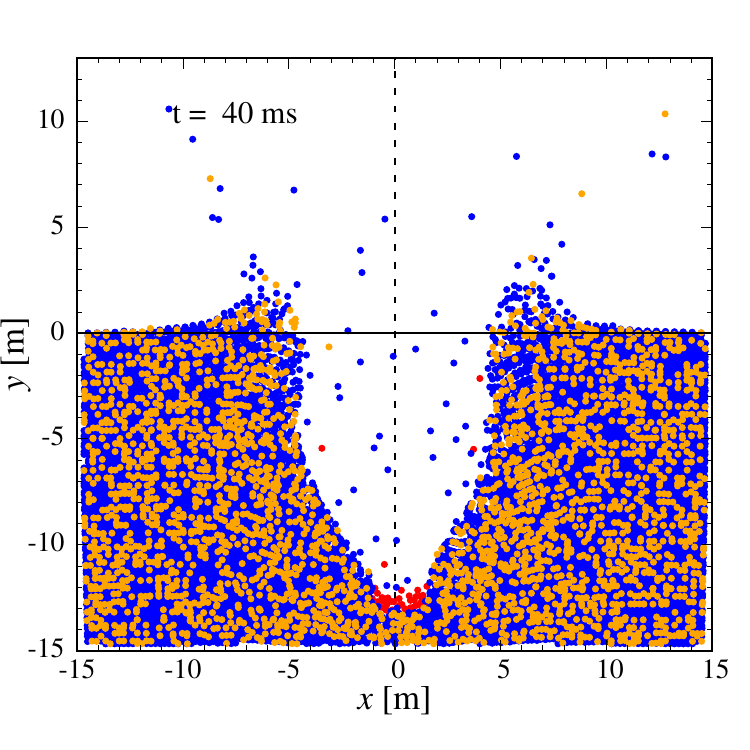}
\includegraphics[scale=0.9]{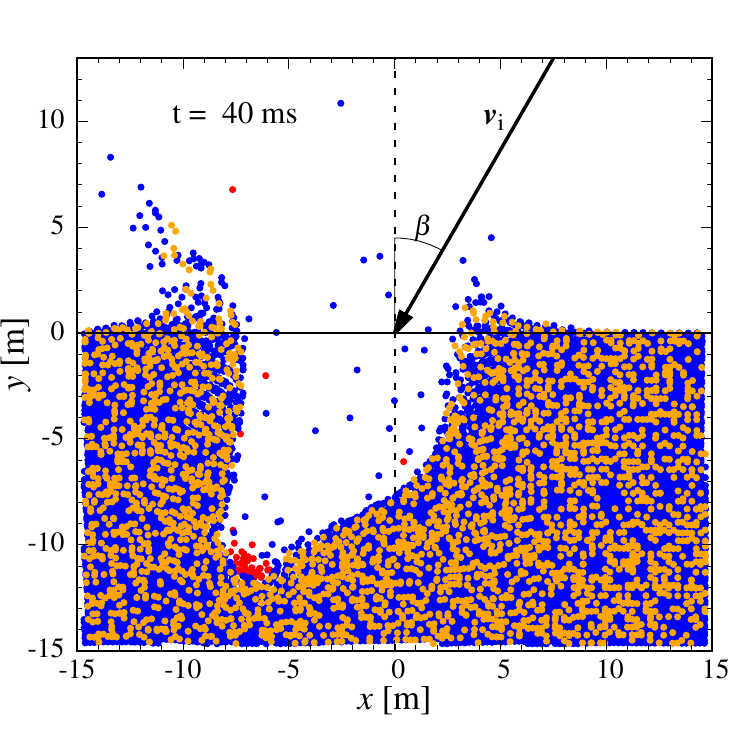}
\vskip -15pt
\includegraphics[scale=0.9]{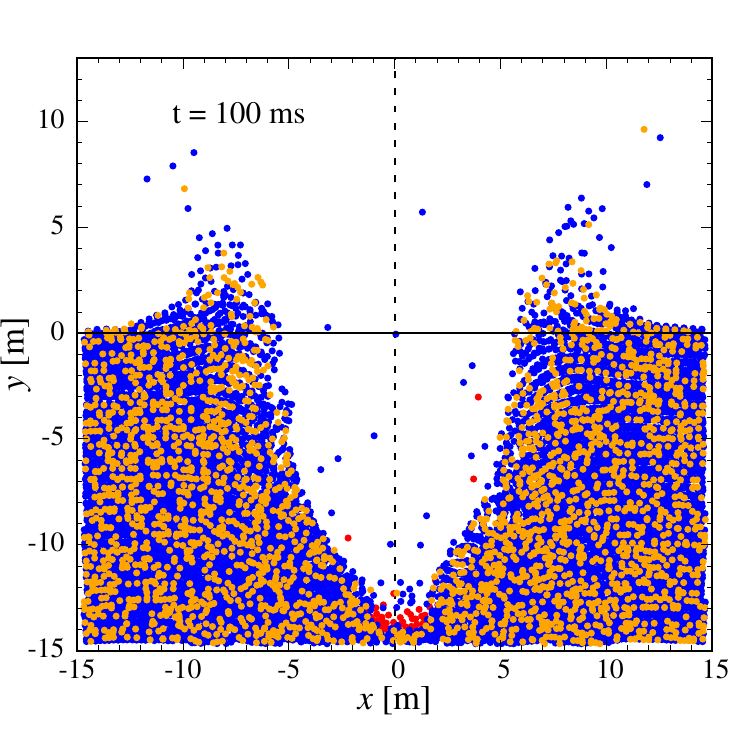}
\includegraphics[scale=0.9]{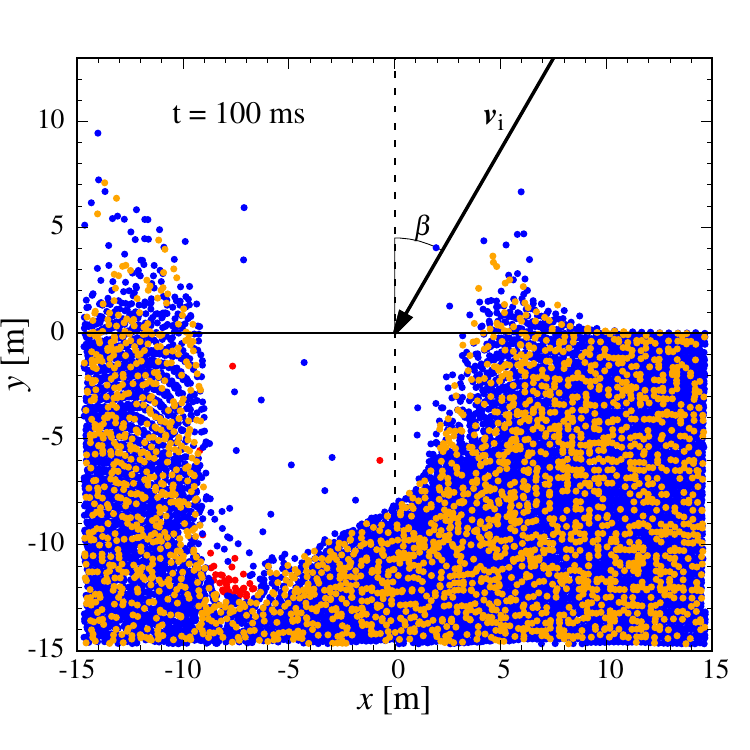}
\vskip -18pt
\caption{Snapshots of the collision of a m-sized object with a porous basaltic target with 
50\% water-mass fraction. The degree of porosity is 50\%. The impactor is pure basalt with 
no porosity (red). The impact velocity is 4.4\,km/s. The impact angle is $\beta=0$ 
(left) and $30^\circ$ (right). The orange color represents porous basalt and blue is for porous ice.
The panels show 2D slices of 3D data.}
\end{figure}

\clearpage

\begin{figure*}
\center
\includegraphics[scale=0.9]{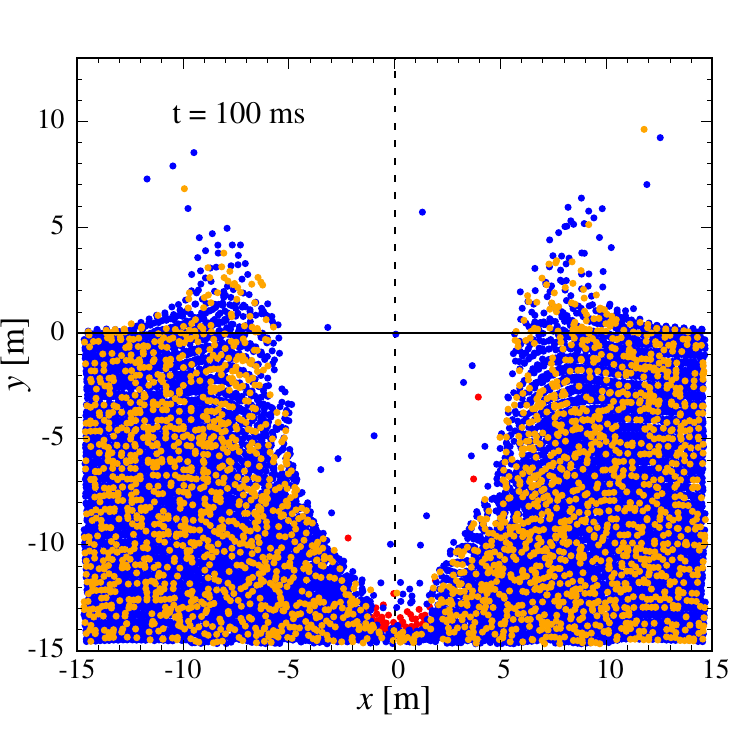}
\includegraphics[scale=0.9]{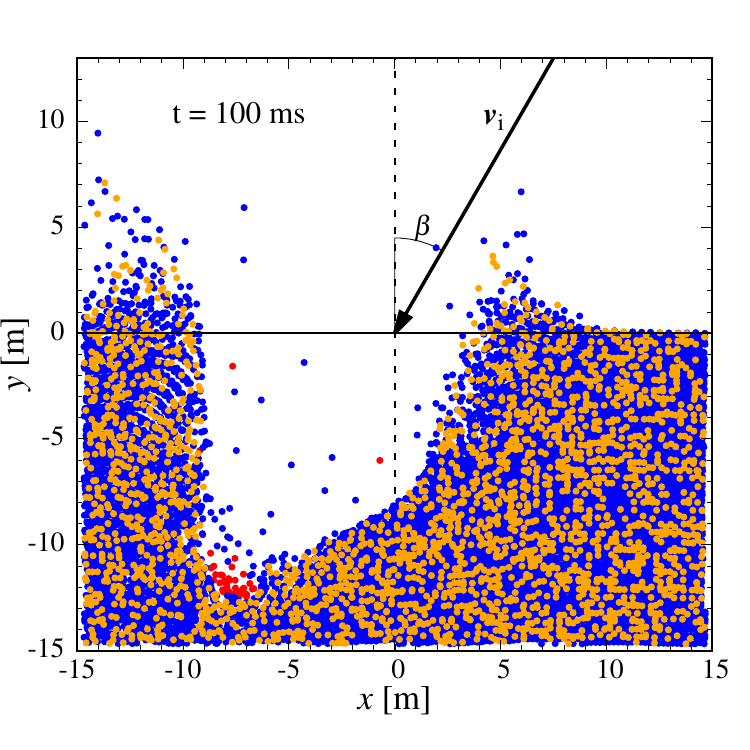}
\vskip -15pt
\includegraphics[scale=0.9]{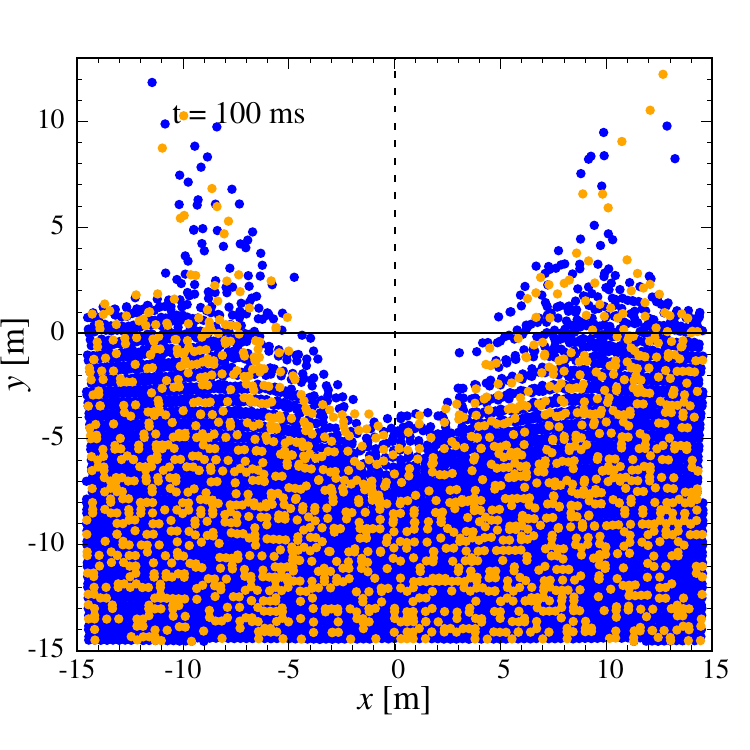}
\includegraphics[scale=0.9]{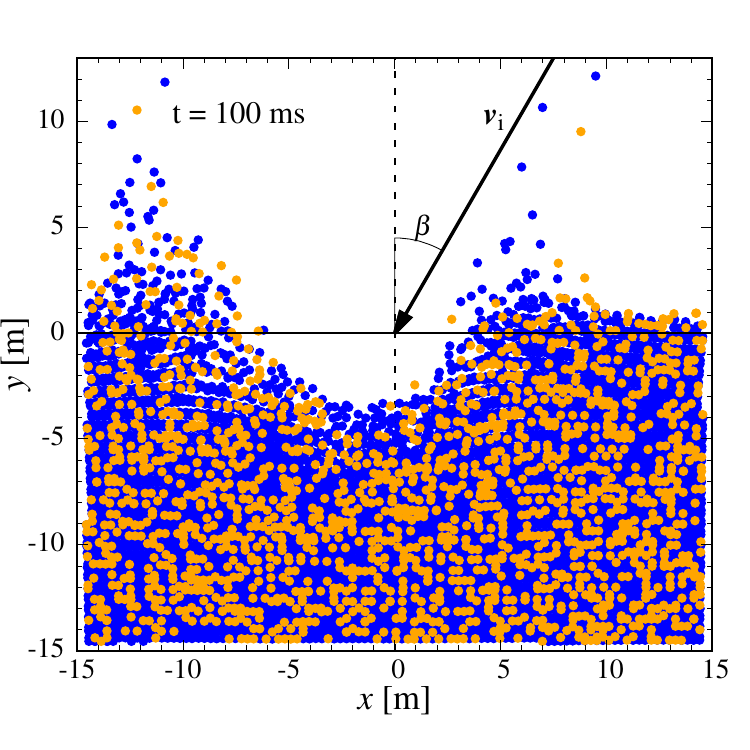}
\vskip -15pt
\caption{Comparing the depths and surface areas of impact craters with and without porosity.
The target in the top panels is porous basalt with 50\% water-mass fraction. The target in 
the bottom panels is non-porous basalt with  50\% water-mass fraction. The impact velocity in
all panels is 4.4 km/s. In the top panels, the orange color represents porous basalt and blue 
is for porous ice. In the bottom panels, the orange color represents non-porous basalt and blue 
is for non-porous ice. The panels show 2D slices of 3D data.}
\end{figure*}

\clearpage

\begin{figure}
\vskip -5in
\center
\includegraphics[scale=0.75]{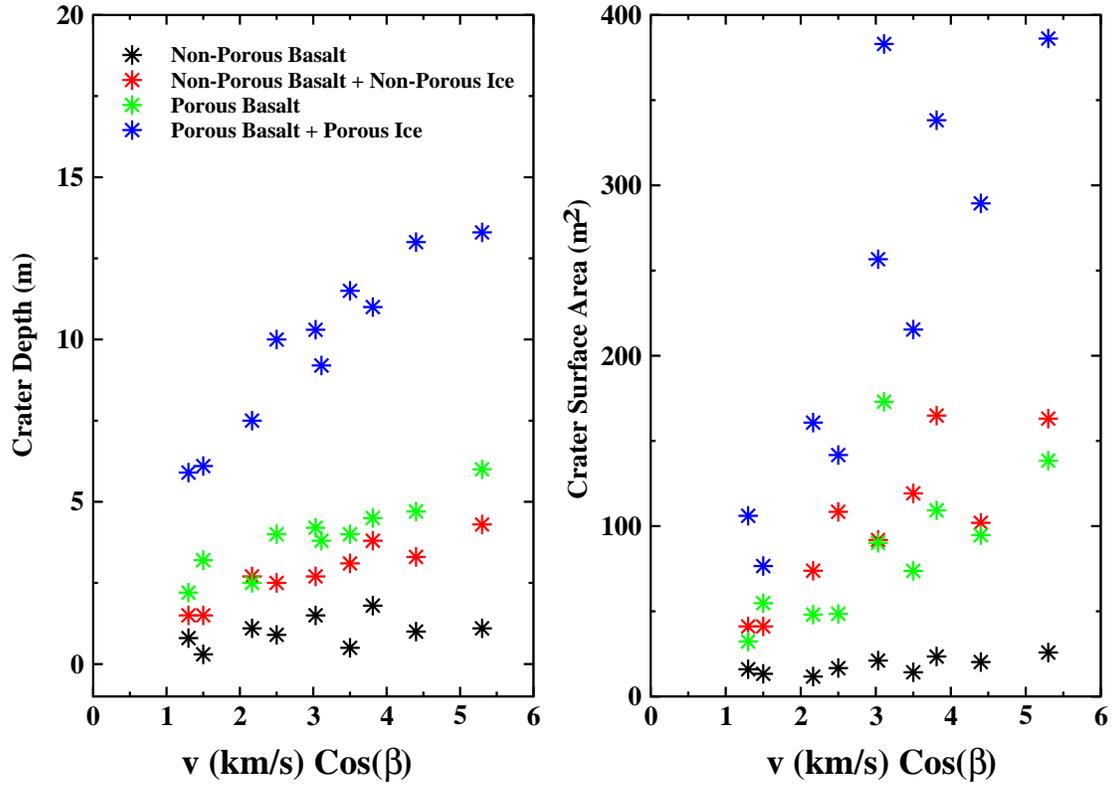}
\vskip 10pt
\caption{Graphs of the depth (left) and surface area (right) of impact craters in terms of
impact velocity for porpous and non-porous targets, and with different water contents.}
\end{figure}

\clearpage

\begin{figure}
\vskip -5in
\center
\includegraphics[scale=0.7]{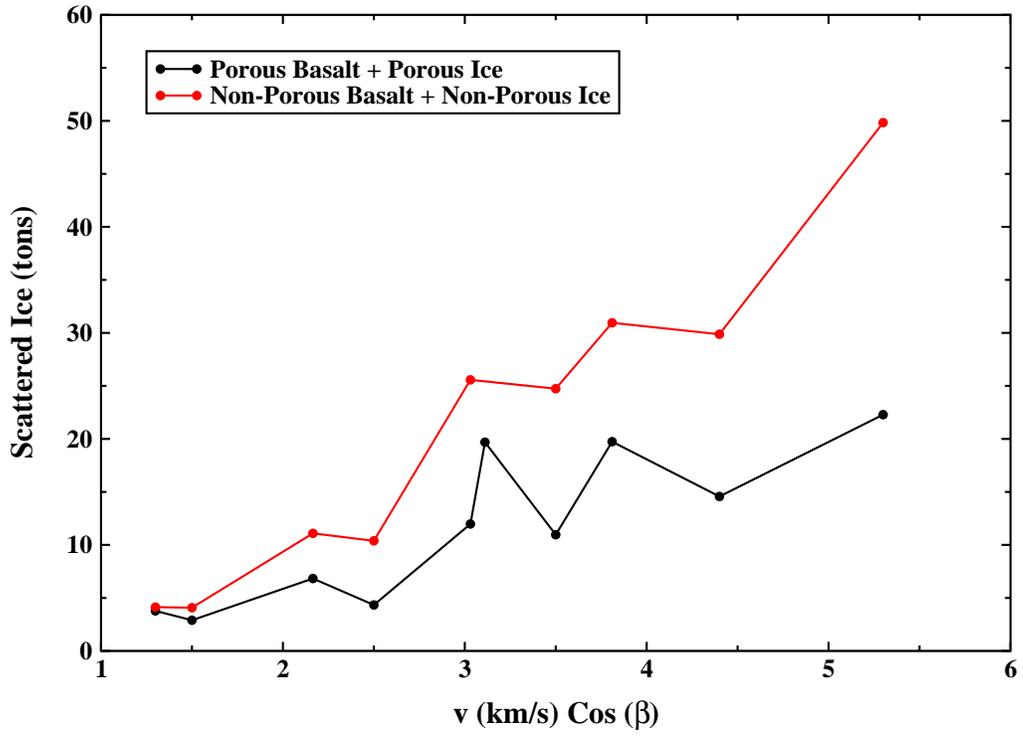}
\caption{Graphs of the scattered ice in terms of impact velocity for porous and
non-porous targets.}
\end{figure}

\clearpage

\begin{figure*}
\center
\includegraphics[scale=1.27]{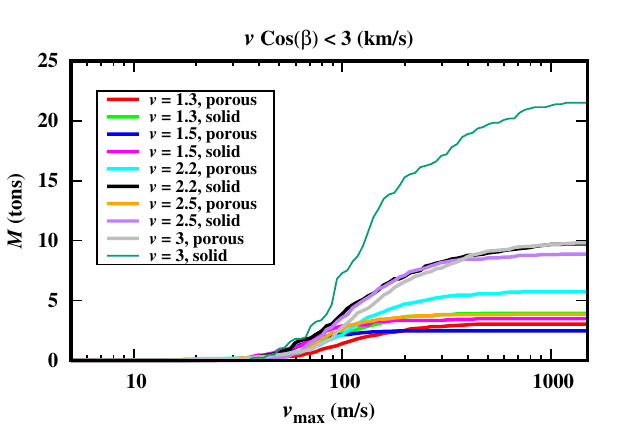}
\includegraphics[scale=1.27]{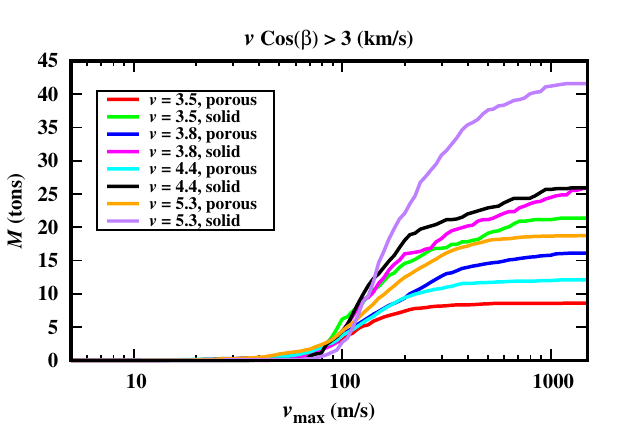}
\includegraphics[scale=1.27]{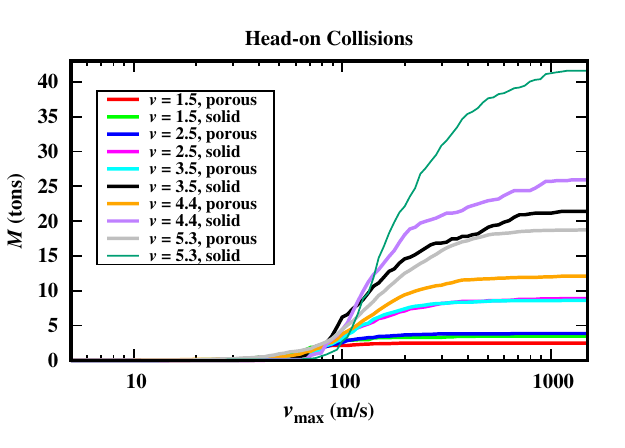}
\includegraphics[scale=1.27]{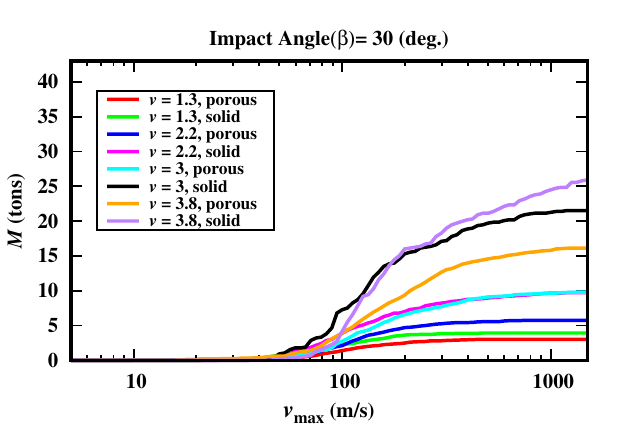}
\caption{Graphs of the accumulative mass of the ejected ice in term of its velocity
for different impact scenarios and impact velocities. The top panel corresponds to
impacts for which the vertical component of the impact velocity is smaller ({\it left})
or larger ({\it right}) than 3 km/s. The bottom panels show ice ejection for a head-on
({\it left}) and a $30^\circ$ ({\it right}) collision. As shown here, in all scenarios,
the velocity of the ejected ice is larger than 20 m/s. The escape velocity of the  currently 
known MBCs is smaller than 2.2 m/s. This figure shows that all ice is ejected and the
amount of re-accreted ice is negligibly small.}

\end{figure*}

\clearpage

\begin{figure*}
\vskip -0.5in
\center
\includegraphics[scale=1]{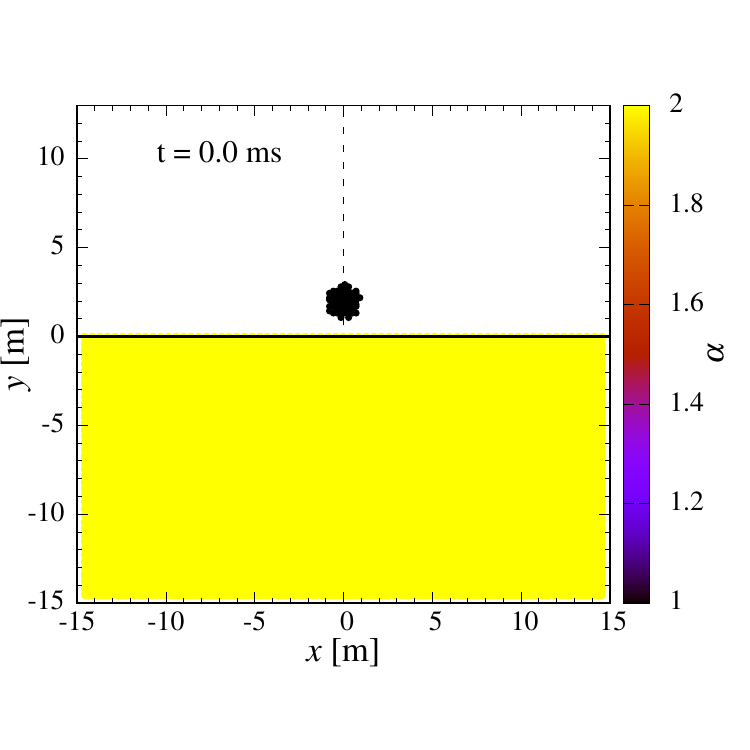}
\includegraphics[scale=1]{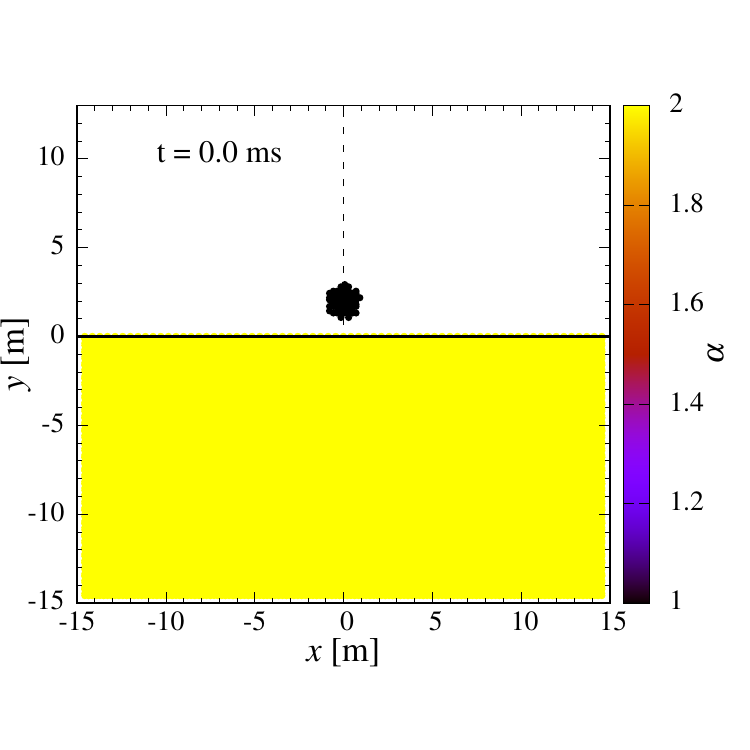}
\vskip -45pt
\includegraphics[scale=1]{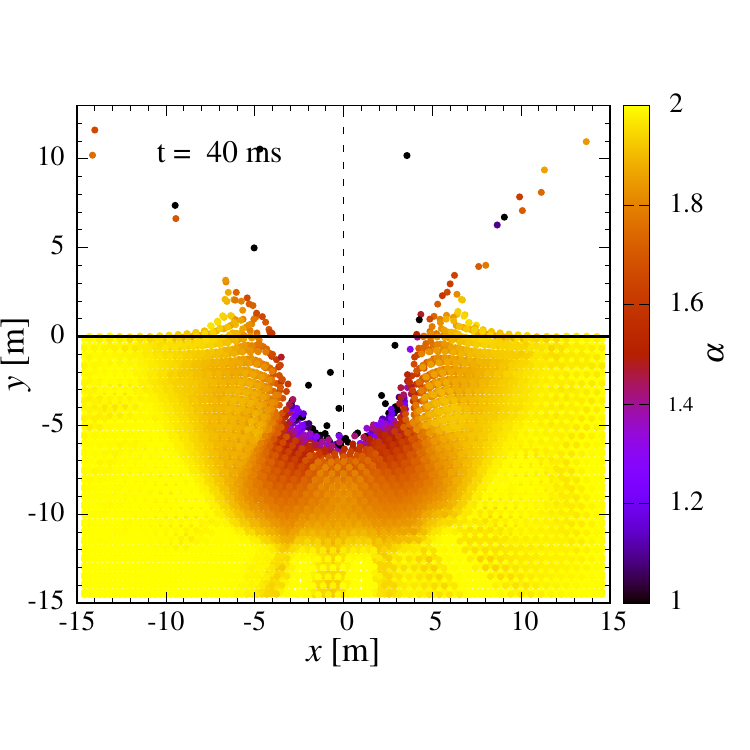}
\includegraphics[scale=1]{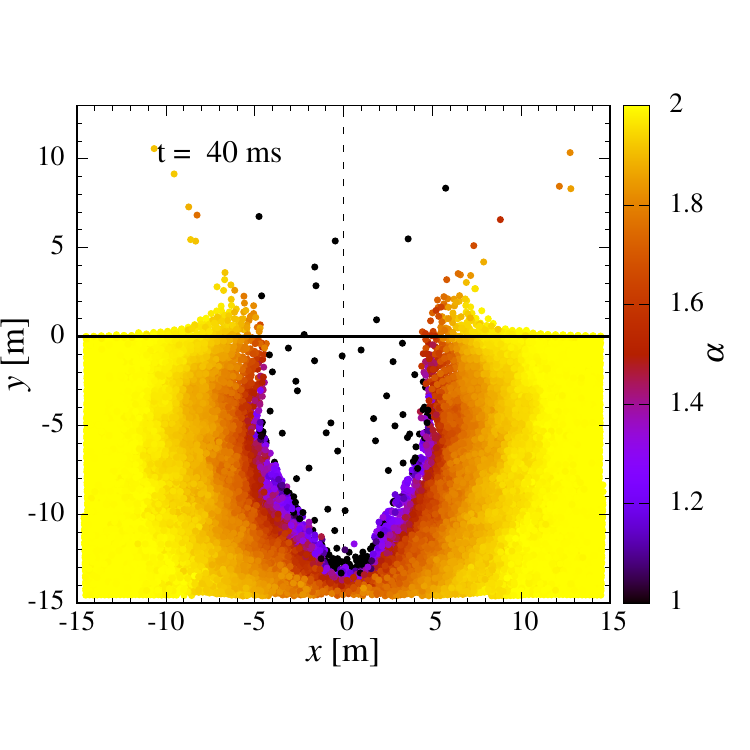}
\vskip -45pt
\includegraphics[scale=1]{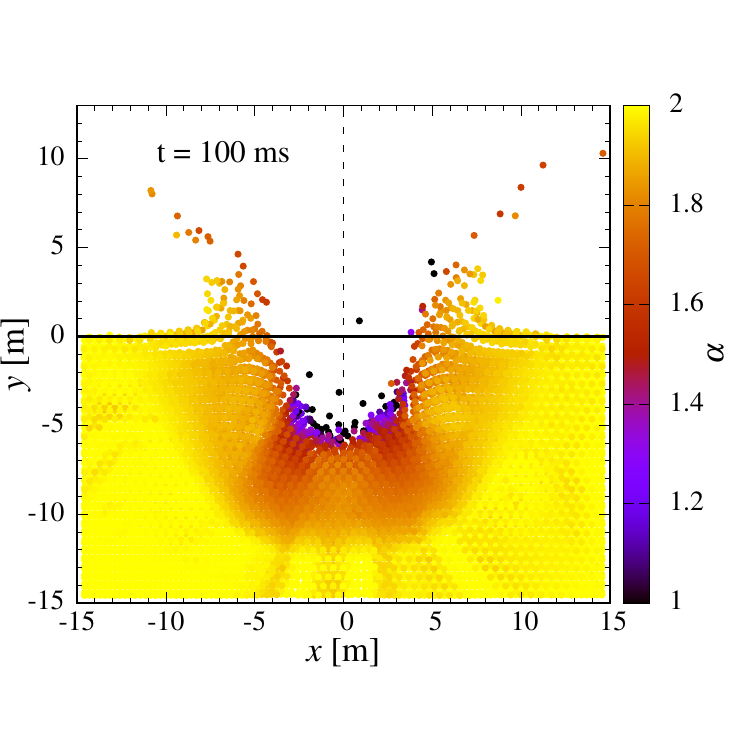}
\includegraphics[scale=1]{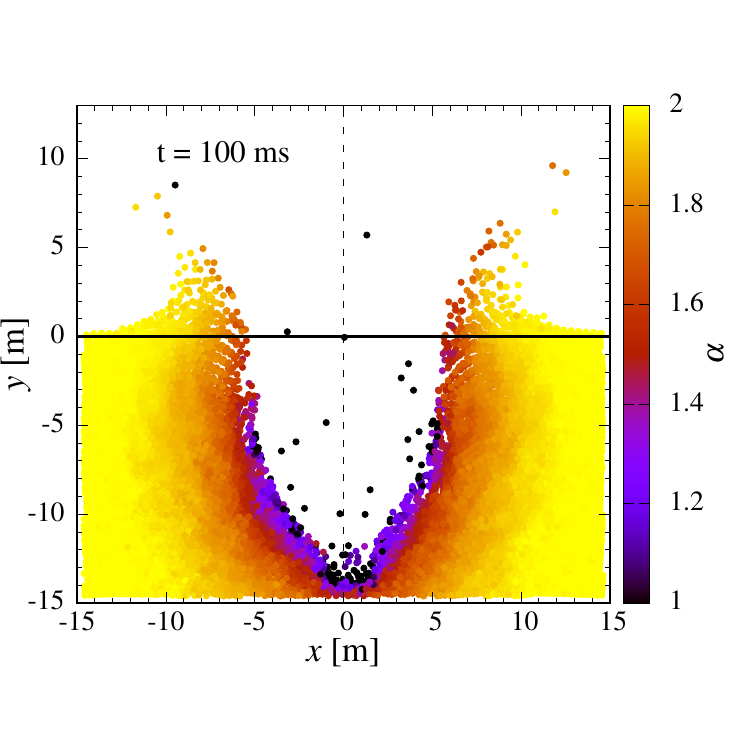}
\vskip -30pt
\caption{Snapshots of the variation of the porosity of the target in figure 1 during an impact. 
The color coding represents the value of the distention parameter $\alpha$ corresponding to the degree of 
compaction and porosity of the object. To better demonstrate changes in porosity, we show in the left 
column a dry, basaltic target with 50\% porosity ($\alpha=2$) and in the right column, we use the same target
but this time with 50\% water-mass fraction of porous ice. As shown here, the object is compacted 
at the site of the impact and the compacation extends to its inner parts
as the shock of the impact propagates inside the body. However, most of the interior part 
of the target maintain its original porosity. The panels show 2D slices of 3D data. }
\end{figure*}

\clearpage

\begin{figure}
\vskip -3in
\center
\includegraphics[scale=0.9]{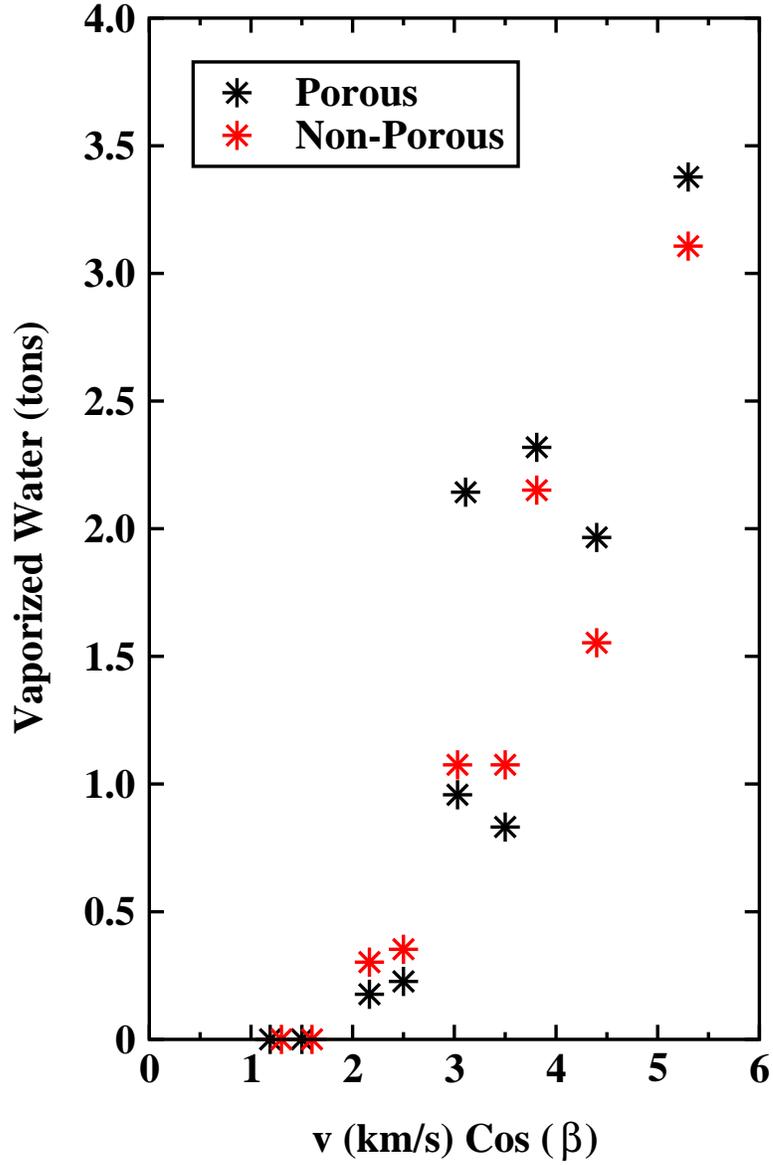}
\vskip -25pt
\caption{Graphs of the vaporized ice due to the heat of impact in terms of the impact velocity for porous 
and non-porous targets.}
\end{figure}

\clearpage

\begin{table*}[ht]
\center
\normalsize
\caption{\normalsize Physical properties of the currently known MBCs\label{t:mbcdata}. 
The quantity $D_{\rm e}$ represents
an MBC's effective diameter, $a\,, e$ and $i$ are its semimajor axis, eccentricity and orbital inclination,
respectively, $T_{\rm J}$ is its Tisserand number with respect to Jupiter, and $v_{\rm esc}$ is the value of its
escape velocity. Adopted from \citet{Jewitt15}.}
\smallskip
\begin{tabular}{lccccccc}
\hline\hline
\multicolumn{1}{c}{Object} &
\multicolumn{1}{c}{$D_\mathrm{e}$ [km]} &
\multicolumn{1}{c}{$a\, \mathrm{[AU]}$} &
\multicolumn{1}{c}{$e$} &
\multicolumn{1}{c}{$i \mathrm{[deg]}$} &
\multicolumn{1}{c}{$T_{\rm J}$} &
\multicolumn{1}{c}{$v_\mathrm{esc}\, \mathrm{[m/s]}\/$} &
\multicolumn{1}{c}{Ref. $^a$} \\
\hline
133P/(7968) Elst-Pizarro  &   $3.8\pm0.6$	&   3.157	&  0.165   &  1.39  &  3.184   &  2.13  &  1  \\
176P/(118401)LINEAR	  &   $4.0\pm0.4$	&   3.196	&  0.192   &  0.24  &  3.167   &  1.95  &  1  \\
238P/Read (P/2005 U1)     &   0.8               &   3.165	&  0.253   &  1.27  &  3.152   &  ....  &  1  \\
259P/Garradd (P/2008 R1)  &   $0.3\pm0.02$      &   2.726	&  0.342   &  15.90 &  3.216   &  0.62  &  2  \\
324P/La Sagra (P/2010 R2) &   1.1       	&   3.099       &  0.154   &  21.39 &  3.100   &  0.49  &  3,4 \\
288P/(300163) 2006 ${\rm VW}_{139}$   & 3        &   3.050       &  0.200   &  3.24  &  3.203   &  ...   &  5 \\
P/2012 T1 (PANSTARRS)     &   2.4               &   3.154       &  0.236   &  11.06 &  3.134   &  ...   &  6  \\
313P/Gibbs (P/2014 S4)    &   1.0	        &   3.156       &  0.242   &  10.97 &  3.132   &  0.86  &  7  \\
\hline\hline
\end{tabular}
\label{table1}
\tablenotetext{a}{1=\citet{Hsieh06}, 2=\citet{Jewitt09}, 3=\citet{Hsieh12a}, 4=\citet{Hsieh15}, 
5=\citet{Hsieh12b}, 6=\citet{Hsieh13}, 7=\citet{Hsieh15}}
\end{table*}

\begin{deluxetable}{lccccccccccccc}
\tabletypesize{\scriptsize}
\rotate
\tablecaption{Material parameters for basalt and ice. The quantity $\varrho_0$ is the bulk density of the object.
The 10 quantities ${\rho_0},\,{A_{\rm T}},\,{B_{\rm T}},\,{E_0},\, {E_{\rm {iv}}},\,{E_{\rm {cv}}},\, {a_{\rm T}},\, 
{b_{\rm T}},\, {\alpha_{\rm T}}$ and 
$\beta_{\rm T}$ are the parameters used in the Tillotson equation of state \citep{Melosh96}. The remaining quantities,
$K,\, \mu$, and $Y_0$ are the bulk modulus, shear modulus, and yield stress, respectively.
Values for basalt and ice are taken from \citep{Benz99}. Note that $A_\mathrm{T}\/$ 
and $B_\mathrm{T}\/$ are set equal to the bulk modulus. }
\tablewidth{0pt}
\tablehead{
\colhead{Material} & \colhead{$\varrho_0$}[$\mathrm{kg/m^{3}}$] & \colhead{$A_\mathrm{T}$ [GPa]} & 
\colhead{$B_\mathrm{T}$ [GPa]} & \colhead{$E_0$ [$\mathrm{MJ/kg}$]} & \colhead{$E_\mathrm{iv}$ [$\mathrm{MJ/kg}$]} & 
\colhead{$E_\mathrm{cv}$ [$\mathrm{MJ/kg}$]} & \colhead{{$a_\mathrm{T}$}} & \colhead{$b_\mathrm{T}$} & 
\colhead{$\alpha_\mathrm{T}$} & \colhead{$\beta_\mathrm{T}$} & \colhead{$K$ [GPa]} & \colhead{$\mu$ [GPa]} & 
\colhead{\tstrut $Y_0$ [GPa]}}
\startdata
Basalt & 2700 & 26.7  & 26.7  & 487 & 4.72 & 18.2  & 0.5 & 1.50 &  5 &  5 & 26.7 & 22.7 & 3.5 \\
Ice & 917 & 9.47 & 9.47 &10 & 0.773 & 3.04 & 0.3 & 0.1 & 10 & 5 & 9.47 & 2.8 & 1 \\
\enddata
\end{deluxetable}


\begin{thebibliography}{}

\bibitem[Asphaug et al. (2002)]{Asphaug02}
Asphaug, E., Ryan, E. V. \& Zuber, M. T. 2002, in: Asteroids III, W. F. Bottke Jr., A. Cellino, 
P. Paolicchi, and R. P. Binzel (eds), University of Arizona Press, Tucson, p.463-484

\bibitem[Basilevsk et al. (2016)]{Basilevsk16}
Basilevsky, A. T., Krasil\'nikov, S. S., Shiryaev, A. A. et al. 2016, Sol. Syst. Res., 50, 225 

\bibitem[Benz \& Asphaug (1994)]{Benz94}
Benz, W. \& Asphaug, E. 1994, Icarus, 107, 98

\bibitem[Benz \& Asphaug (1999)]{Benz99}
Benz, W. \& Asphaug, E. 1999, Icarus, 142, 5

\bibitem[Bottke et al. (1994)]{Bottke94}
Bottke, W. F., Nolan, M. C., Greenberg, R. \& Kovoord, R. A. 1994, Icarus, 107, 255

\bibitem[Carroll \& Holt (1972)]{Carroll72}
Carroll, M. M. \& Holt A. C. 1972, J. Appl. Phys., 43, 1626

\bibitem[Collins et al. (2009)]{Collins09}
Collins, G. S., Davison, T., Elbeshausen, D. \&  W\"unnemann, K.
2009, LPSC contribution 1620

\bibitem[Collins (2014)]{Collins14}
Collins, G. S. 2014, JGR: Planet, 119, 2600

\bibitem[Grady \& Kipp (1993)]{Grady93}
Grady, D. E. \& Kipp, M. E. 1993, in: J. R. Asay \& M. Shahinpoor (Eds), 
High Pressure Shock Compression of Solids (Springer-Verlag, New York), Chapter 5, 265

\bibitem[Haghighipour (2009)]{Hagh09}
Haghighipour, N. 2009, Meteor. \& Planet. Sci., 44, 1863

\bibitem[Haghighipour (2010)]{Hagh10}
Haghighipour, N. 2010, in: Icy Bodies of the Solar System, Proceedings of the IAU Symposium 263, 207

\bibitem[Haghighipour et al. (2016)]{Hagh16}
Haghighipour, N., Maindl, T. I., Sch\"afer, C., Speith, R. \& Dvorak, R. 2016, \apj, 830, 22

\bibitem[Herrmann (1969)]{Herrmann69}
Herrmann, W. 1969, J. Appl. Phys., 40, 2490

\bibitem[Holsapple (1993)]{Holsapple93}
Holsapple K. A. 1993, Annu. Rev. Earth Planet. Sci., 21, 333

\bibitem[Hsieh et al. (2004)]{Hsieh04}
Hsieh, H. H., Jewitt, D. C. \& Fern\'andez, Y. R. 2004, \aj, 127, 2997

\bibitem[Hsieh \& Jewitt (2006)]{Hsieh06}
Hsieh, H.~H. \& Jewitt, D. 2006, Science, 312, 561

\bibitem[Hsieh et al. (2012a)]{Hsieh12a}
Hsieh H. H., Yang, B., Haghighipour, N., Novakovi\'c, B., et al. 2012a, \aj, 143, 104

\bibitem[Hsieh et al. (2012b)]{Hsieh12b}
Hsieh H. H., Yang, B., Haghighipour, N., Kaluna, H. M., et al. 2012b, \apj, 748, L15

\bibitem[Hsieh et al. (2013)]{Hsieh13}
Hsieh H. H., Kaluna, H. M., Yang, B., Novakovi\'c, B., et al. 2013, \apj,771, L1  

\bibitem[Hsieh et al. (2015)]{Hsieh15}
Hsieh H. H., Hainaut, O., Novakovi\'c, B., Bolin, B., et al. 2015, \apj, 800, L16

\bibitem[Hsieh \& Haghighipour (2016)]{Hsieh16}
Hsieh, H. H. \& Haghighipour, N. 2016, \icarus, 277, 19

\bibitem[Graves et al. (2017)]{Graves17}
Graves, K., Minton, D. A., Molaro, J. \& Hirabayashi, M. 2017, American Astronomical Society, 
DPS meeting 49, Abstract id.100.05

\bibitem[Jewitt et al. (2009)]{Jewitt09}	
Jewitt, D., Yang, B. \& Haghighipour, N. 2009, \apj, 137, 4313

\bibitem[Jewitt et al. (2015)]{Jewitt15}
Jewitt, D., Hsieh, H. \& Agarwal, J. 2015, in P. Michel, F. DeMeo \& W. Bottke (Eds),
ASTEROIDS IV (University of Arizona Press, Tucson), 221

\bibitem[Jutzi et al. (2008)]{Jutzi08}
Jutzi, M., Benz, W. \& Michel, P. 2008, \icarus, 198, 242

\bibitem[Jutzi et al. (2009)]{Jutzi09}
Jutzi, M., Michel, P., Hiraoka, K., Nakamura, A. M. \& Benz, W. 2009, \icarus, 201, 802

\bibitem[Maindl et al. (2013)]{Maindl13}
Maindl, T. I., Sch\"afer, C., Speith, R., S\"uli, \'A., Forg\'acs-Dajka, E. \& Dvorak, R. 2013,
AN, 334, 996

\bibitem[Maindl et al. (2014)]{Maindl14}
Maindl, T. I., Dvorak, R., Sch\"afer, C. \& Speith, R. 2014 , in: Kne\v zevi\'c, Z. \& Lema\^itre, A. (Eds),
Complex Planetary Systems, Proceedings of the IAU Symposium 310, 138

\bibitem[Maindl et al. (2015)]{Maindl15}
Maindl, T. I., Dvorak, R., Lammer, H., G\"udel, M., Sch\"afer, C., Speith, R., Odert, P.,
Erkaev, N. V., Kislyakova, K. G. \& Pilat-Lohinger, E. 2015, A\&A, 574, A22 

\bibitem[Melosh (1996)]{Melosh96}
Melosh, H. J. 1996, Impact Cratering, Oxford University Press

\bibitem[Milbury et al. (2015)]{Milbury15}
Milbury, C., Johnson, B. C., Melosh, H. J., Collins, G. S., Blair, D. M., 
Soderblom, J. M., Nimmo, F., Bierson, C. J., Phillips, R. J. \& Zuber, M. T.
2015, GeoRL, 42, 9711

\bibitem[Nesvorn\'y et al. (2008)]{Nesvorny08}
Nesvorn\'y, D., Bottke, W. F., Vokrouhlick\'y, D., et al. 2008, \apj, 679, 143

\bibitem[Prialnik \& Rosenberg (2009)]{Prialnik09}
Prialnik, D. \& Rosenberg, E.~D. 2009, \mnras, 399, L79

\bibitem[Richardson et al. (2007)]{Richardson07}
Richardson J. E., Melosh H. J., Lisse C. M. \& Carcich B. 2007, Icarus, 190, 357

\bibitem[Sch\"afer et al. (2007)]{Schafer07}
Sch\"afer, C., Speith, R. \& Kley, W. 2007, A\&A, 470, 733

\bibitem[Sch\"afer et al. (2016)]{Schafer16}
Sch\"afer, C., Riecker, S., Maindl, T. I., Scherrer, S., Speith, R. \& Kley, W. 2016, A\&A, 590, A19

\bibitem[Sch\"orghofer (2008)]{Schorghofer08}
Sch\"orghofer, N. 2008, \apj, 682, 697

\bibitem[Sheppard \& Trujillo (2015)]{Sheppard15}
Sheppard, S. S., Trujillo, C. 2015, AJ, 149, article id. 44

\bibitem[Steckloff et al. (2016)]{Steckloff16}
Steckloff, J. K., Graves, K., Hirabayashi, M., Melsoh, H. J. \& Richardson, J. E. 2016, Icarus, 272, 60

\bibitem[Tillotson (1962)]{Tillotson62}
Tillotson, J. H. 1962. Metallic Equations of State for Hyper-velocity Impact. General Atomic Report GA-3216, 
General Atomic, San Diego, CA.

\end{thebibliography}
\end{document}